\documentclass[prl,twocolumn,showpacs,showkeys,nofootinbib,superscriptaddress]{revtex4}
\usepackage{graphicx}
\usepackage{dcolumn}
\usepackage{bm}

\def\sss{\mbox{\boldmath $\sigma$}}
\def\tt{\mbox{\boldmath $\tau$}}

\newcommand{\be}{\begin{equation}}
\newcommand{\ee}{\end{equation}}
\newcommand{\bea}{\begin{eqnarray}}
\newcommand{\eea}{\end{eqnarray}}

\begin{document}

\topmargin -0.50in

\title{Can one measure nuclear matrix elements of neutrinoless double beta decay?}

\author{Vadim Rodin}
\email{vadim.rodin@uni-tuebingen.de}
\affiliation{Institut f\"{u}r Theoretische Physik der Universit\"{a}t
T\"{u}bingen, D-72076 T\"{u}bingen, Germany}
\author{Amand Faessler}
\affiliation{Institut f\"{u}r Theoretische Physik der Universit\"{a}t
T\"{u}bingen, D-72076 T\"{u}bingen, Germany}
\begin{abstract}

By making use of the isospin conservation by strong interaction,
the Fermi $0\nu\beta\beta$ nuclear matrix element $M_{F}^{0\nu}$ is 
transformed to acquire the form of an energy-weighted double Fermi transition matrix element. 
This useful representation allows reconstruction of the total $M_{F}^{0\nu}$
provided a small isospin-breaking Fermi matrix element between the isobaric analog state 
in the intermediate nucleus and the ground state of the daughter nucleus could be measured, 
e.g., by charge-exchange reactions. 
Such a measurement could set a scale for the $0\nu\beta\beta$ nuclear matrix elements 
and help to discriminate between the different nuclear structure models in which calculated 
$M_{F}^{0\nu}$  may differ by as much as a factor of 5 
(that translates to about 20\% difference in the total $M^{0\nu}$).

\end{abstract}



\date{\today}
\maketitle

Neutrino is the only known spin-1/2 fermion which may be truly neutral, i.e., identical with its own antiparticle. In such a case one speaks about Majorana neutrino, to be contrasted with Dirac neutrino which is different from its antiparticle~\cite{Kay89,Moh91}. 
Majorana neutrinos naturally appear in many extensions of the standard model (see, e.g.,~\cite{Moh04}). 
Also, the smallness of neutrino masses (more than five orders of magnitude 
smaller than the electron mass) finds an elegant explanation within the see-saw model which assumes neutrinos to be Majorana particles~\cite{see-saw}.

The fact that neutrinos have mass has firmly been established by neutrino oscillation experiments 
(for reviews see, e.g., Ref.~\cite{Kay08}).
However, the observed oscillations cannot in principle pin down the absolute scale of the neutrino masses.
A study of nuclear neutrinoless double beta ($0\nu\beta\beta$) decay~~$^A_Z {\mathrm{El}}_N \longrightarrow \ _{Z+2}^{\phantom{+2} A} {\mathrm{El}}_{N-2} + 2e^-$ 
offers a mean to probe the absolute neutrino masses at the level of tens of meV. 

Double beta decay is a rare decay process which may occur in the second order of weak interaction. It 
offers the only feasible way to test the charge-conjugation property of neutrinos. 
The existence of the $0\nu\beta\beta$ decay would immediately prove neutrino to be a Majorana particle as 
assured by the Schechter-Valle theorem~\cite{SV82}. The $0\nu\beta\beta$ decay is strictly forbidden in the standard model of the electroweak interaction in which the lepton number is  conserved, thus its observation would be of paramount importance for our understanding of 
particle physics beyond the standard model~\cite{vogelbook,fae98,AEE07}.

The next generation of $0\nu\beta\beta$-decay experiments (CUORE, GERDA, MAJORANA, SuperNEMO etc., 
see, e.g., Ref.~\cite{AEE07}  for a recent review) has a great discovery potential. 
Provided the corresponding decay lifetimes are accurately measured, 
knowledge of the relevant nuclear matrix elements (m.e.) $M^{0\nu}$ will become indispensable 
to reliably deduce the effective Majorana mass 
from the experimental data. 

Two basic theoretical approaches are used to evaluate $M^{0\nu}$, the quasiparticle random phase approximation 
(QRPA)~\cite{Rod03a,Kort07}, 
including its continuum version~\cite{Rod08}, and the nuclear shell model (NSM)~\cite{cau96}.
There has been great progress in the calculations over the last five years, and now the QRPA $0\nu\beta\beta$ nuclear m.e. of different groups seem to converge. 
However, the NSM $M^{0\nu}$ are systematically and substantially (up to a factor of 2) 
smaller than the corresponding QRPA ones. There is now an active discussion in literature on what could be the reason of such a discrepancy, a too small single-particle model space of the NSM or a neglect of complex nuclear configurations within the QRPA.
Even more striking is the difference in the Fermi contribution to the total $M^{0\nu}$ which can be up to a factor of 5 larger in the QRPA calculations than in the NSM ones.

In view of this situation, 
it would be extremely important to find a possibility to determine $M^{0\nu}$ experimentally.
There have been attempts to reconstruct the nuclear amplitude of two-neutrino $\beta\beta$ decay 
(which experimentally is very accurately known from the direct counting $\beta\beta$-decay experiments~\cite{AEE07}) 
from partial one-leg transition amplitudes 
to the intermediate $1^+$ states measured in charge-exchange reactions~\cite{frekers}. 
However, 
such a procedure can consistently determine $M^{2\nu}$ only if a transition via a single intermediate 
$1^+$ state dominates $M^{2\nu}$ (the so-called single-state dominance). 
In the case of comparable contributions of several intermediate $1^+$ states the results 
from charge-exchange reactions cannot directly provide $M^{2\nu}$,
since relative phases of the contributions cannot be measured. 
Pursuing the same way to reconstruct $M^{0\nu}$ seems even more hopeless, since  
many intermediate states of all multipolarities (with a rather moderate contribution of the $1^+$ states)
are virtually populated in the $0\nu\beta\beta$ decay due to a large momentum of the virtual neutrino.

The aim of this Rapid Communication is to suggest a way by which at least the Fermi component of $M^{0\nu}$ can directly 
be measured, e.g., in charge-exchange reactions. For the derivation of the master 
expressions~(\ref{MFtot}),(\ref{MFappr}) below the well-known property of 
the Coulomb interaction to be the leading source of the isospin breaking in nuclei 
is exploited~\cite{auer72,auer83}. Such a measurement of $M^{0\nu}_F$ could set a scale 
for the $0\nu\beta\beta$ nuclear m.e. and help to discriminate between  
different nuclear structure models in which calculated $M_{F}^{0\nu}$ may differ by as much as a factor of 5.


We start our derivation by writing down the $0\nu\beta\beta$ nuclear m.e. in the closure approximation in which it acquires the form $M^{0\nu}=\langle 0_f | \hat W^{0\nu} |0_i\rangle$ of the m.e. of a two-body scalar operator
$\hat W^{0\nu}$ between the parent and daughter ground states $|0_i\rangle$ and $|0_f\rangle$, respectively. 
\footnote{Using closure of the states of the intermediate nucleus $_{Z+1}^{\phantom{+2} A} {\mathrm{El}}_{N-1}$ which are virtually excited in $\beta\beta$-decay would be an exact procedure if there were no energy dependence in the $0\nu\beta\beta$ transition operator. A weak energy dependence of the operator leads in reality to a ``beyond-closure'' correction to the total $M^{0\nu}$ of less than 10\%.}
The total $0\nu\beta\beta$-decay operator $\hat W^{0\nu}\equiv g_A^2\hat W^{0\nu}_{GT}- g_V^2\hat W^{0\nu}_F$ is the sum of the Gamow-Teller and Fermi transition operators~\cite{vogelbook}:
\bea
\displaystyle\hat W^{0\nu}&=&\sum_{ab} P_\nu(r_{ab}) \left(g_A^2\sss_a \cdot \sss_b - g_V^2\right)\tau_a^{-} \tau_b^{-}.
\label{nuPot}
\eea
Here, the vector and axial vector coupling constants are $g_V = 1$ and $g_A=1.25$, respectively, and $P_\nu(r_{ab}\equiv|{\vec r}_a-{\vec r}_b|)$ is the 
neutrino potential which in the simplest Coulomb approximation is just reciprocal of the distance between the nucleons:
$P_\nu(r_{ab})=\frac{1}{r_{ab}}$ (for the sake of simplicity we have taken out the nuclear radius $R$
from the usual definition of $P_\nu$~\cite{vogelbook}).
In this approximation 
\be
\displaystyle \hat W^{0\nu}_F = \sum_{ab} P_\nu(r_{ab}) \tau_a^{-} \tau_b^{-} = 
\frac{1}{e^2} \left[\hat T^{-},[\hat T^{-}, \hat V_C] \right],
\label{nuPotF}
\ee
where $\hat T^{-}=\sum_{a}\tau_a^{-}$ is the isospin lowering operator, and $\hat V_C=\displaystyle\frac{e^2}{8}\sum_{a\neq b}\frac{(1-\tau_a^{(3)})(1-\tau_b^{(3)})}{r_{ab}}$ is the operator of Coulomb interaction between protons. Actually, only the isotensor component of the Coulomb interaction 
$\hat V_C^t = \displaystyle\frac{e^2}{8}\sum_{a\neq b} \frac{T^{(2)}_{ab}}{r_{ab}}$, with $T^{(2)}_{ab}\equiv\tau_a^{(3)}\tau_b^{(3)} - \frac{\tt_a\tt_b}3$, 
survives in the double commutator~(\ref{nuPotF}). 
This isotensor Coulomb interaction does contribute to the mean Coulomb field in the nucleus, but it is easy to see that any  mean-field single-particle operator drops out of the double commutator~(\ref{nuPotF}). 
Thus, the expression~(\ref{nuPotF}) is essentially determined by the residual (after separating out the mean-field contribution) two-body isotensor Coulomb interaction. 

The total nonrelativistic nuclear Hamiltonian $\hat H_{tot}$ consists of the total kinetic energy of nucleons and the strong and Coulomb two-body interactions between them: 
$\hat H_{tot} = \hat T + \hat H_{str} + \hat V_C$. Assuming $\hat H_{str}$ to be exactly isospin-symmetric 
$\left[\hat T^{-}, \hat H_{str}  \right]=0$ 
(we shall quantify later the accuracy of this assertion
but it is well known that the isospin-breaking terms in $\hat H_{str}$ are in fact fairly small~\cite{auer72,auer83}),
one has 
\be
\displaystyle \hat W^{0\nu}_F = \frac{1}{e^2} \left[\hat T^{-},[ \hat T^{-}, \hat H_{tot}] \right],
\label{2comTot}
\ee
and, correspondingly~\cite{Rodin05},
\be
M^{0\nu}_F = - \frac{2}{e^2} 
\sum_s \bar\omega_s \langle 0_f | \hat T^{-} |0^+_s \rangle  \langle 0^+_s | \hat T^{-} |0_i\rangle.
\label{MFtot}
\ee
Here, 
the sum runs over all $0^+$ states of the intermediate $(N-1,Z+1)$ isobaric nucleus, 
$\bar\omega_s=E_s-(E_{0_i}+E_{0_f})/2$
represents the excitation energy of the $s$'th intermediate $0^+$ state relative to the mean energy of the ground states of the initial and final nuclei. To account for the isospin-breaking part of $\hat H_{str}$, $\delta M^{0\nu}_F = \displaystyle \frac{1}{e^2}\langle 0_f |\left[\hat T^{-},[\hat T^{-}, \hat H_{str} ] \right]|0_i\rangle$ should be subtracted from r.h.s. of Eq.~(\ref{MFtot}). 

Among all the intermediate $0^+$ states, the isobaric analog state (IAS) dominates the sum~(\ref{MFtot}). 
In fact, $\langle IAS | \hat T^{-} | 0_i \rangle \approx \sqrt{N-Z}$ is the largest first-leg transition m.e. 
(a few percents of the total Fermi strength $N-Z$ may go to
the highly-excited isovector monopole resonance (IVMR) since the IAS and IVMR get mixed mainly by the Coulomb mean field). 
Similarly, 
the second-leg Fermi transition dominantly populates the 
double IAS (DIAS) in the final nucleus. Due to the isotensor part of the Coulomb interaction (which also gives the only contribution to the double commutator~(\ref{nuPotF})), the final g.s. gets an admixture of the DIAS where the corresponding mixing m.e. is $\displaystyle\langle 0_f | DIAS\rangle=-\frac{\langle 0_f |\hat V_C^t| DIAS\rangle}{E_{DIAS}}$, with $E_{DIAS}\approx 2\bar\omega_{IAS}$. Thereby, one gets $\langle 0_f | \hat T^{-} |IAS \rangle \neq 0$.

Other quantitative arguments for the dominance of the IAS in the sum~(\ref{MFtot}) follow from the representation of the double commutator: 
\bea & 
\displaystyle 
\left[\hat T^{-},[ \hat T^{-}, \hat V_C^t ] \right]= 
\hat V_C^t \left(\hat T^{-}\right)^2 + \left(\hat T^{-}\right)^2 \hat V_C^t - 2 \hat T^{-} \hat V_C^t \hat T^{-}.
\label{x} \nonumber
\eea
It is clear that the first term $V_C^t \left(T^{-}\right)^2$ dominates the m.e. 
$\langle 0_f | \left[\hat T^{-},[ \hat T^{-}, \hat V_C^t ] \right]|0_i\rangle $, since the other 
m.e., because of $\hat T^{+}|0_f\rangle \approx 0$ 
(with a small deviation from zero originating from an isospin symmetry violation effect, caused mainly by the Coulomb mean field), contain additional suppression as compared with the leading term $\langle 0_f |\hat V_C^t \left(\hat T^{-}\right)^2|0_i\rangle=\langle 0_f |\hat V_C^t| DIAS\rangle \langle DIAS |\left(\hat T^{-}\right)^2|0_i\rangle$. 

%
%

Thus, $M^{0\nu}_F$ is determined by the amplitude of the double Fermi transition via the IAS in the intermediate nucleus into the ground state of the final nucleus where the second Fermi transition amplitude is due to an admixture of the DIAS in the final nucleus to the ground state of the parent nucleus: $\langle 0_f |\hat T^{-}| IAS\rangle \langle IAS |\hat T^{-}|0_i\rangle=\langle 0_f | DIAS\rangle \langle DIAS |\hat T^{-}| IAS\rangle \langle IAS |\hat T^{-}|0_i\rangle$.
Finally, one can write
\be
M^{0\nu}_F \approx - \frac{2}{e^2}\,\bar\omega_{IAS} 
\langle 0_f | \hat T^{-} |IAS \rangle  \langle IAS | \hat T^{-} |0_i\rangle .
\label{MFappr}
\ee

Therefore, the total $M^{0\nu}_F$  
can be reconstructed according to Eq.~(\ref{MFappr}), if one is able to measure the $\Delta T=2$ isospin-forbidden m.e. 
$\langle 0_f | \hat T^{-} | IAS \rangle$, for instance in charge-exchange reactions of the $(n,p)$-type 
(also the same m.e. determines $M^{2\nu}_F$, but it would be much more difficult to extract it). 
Using the QRPA calculation results for $M^{0\nu}_F$~\cite{Rod03a,Kort07}, this m.e. can roughly be estimated as 
$\langle 0_f | \hat T^{-} | IAS \rangle \sim 0.005$,  
i.e. about a thousand times smaller than the first-leg m.e. $\langle IAS | \hat T^{-} | 0_i \rangle \approx \sqrt{N-Z}$. 
This strong suppression of $\langle 0_f | \hat T^{-} | IAS \rangle$ reflects the smallness of the isospin violation in nuclei. 
The IAS has been observed as a prominent and extremely narrow resonance and its various features have well been studied 
by means of (p,n), ($^3$He,t) and other charge-exchange reactions, see, e.g.~\cite{Rob95}.
This gives us hope that a measurement of $\langle 0_f | \hat T^{-} | IAS \rangle$ in the $(n,p)$ charge-exchange channel might be possible. More generally, a measurement by whichever experimental mean of the $\Delta T=2$ admixture of the DIAS in the final ground state would be enough to determine $M_{F}^{0\nu}$.

A qualitative analysis of the physics involved in calculations of $M_{F}^{0\nu}$ can be conducted further. One can define an operator $\hat{\cal V}_C^t = \displaystyle\frac{e^2}{8\bar R}\sum_{a b} T^{(2)}_{ab}$ which is obtained by the substitution of $\displaystyle\frac{1}{r_{ab}}$ by a constant $\displaystyle\frac{1}{\bar R}$ in the definition of the isotensor Coulomb interaction.  
Such an operator $\hat{\cal V}_C^t$ is diagonal in the basis of isospin eigenstates and 
does not mix in the first order the DIAS and the final ground state, $\langle DIAS | \hat{\cal V}_C^t|0^+_f \rangle=0$. 
The matrix element 
$\displaystyle\frac{1}{e^2}\langle 0_f |\left[\hat T^{-},[ \hat T^{-},\hat{\cal V}_C^t ] \right]|0_i\rangle
=\frac{1}{2\bar R}\sum_s \langle 0_f | \hat T^{-} |0^+_s \rangle  \langle 0^+_s | \hat T^{-} |0_i\rangle$ 
is by a large factor $\displaystyle\frac{e^2}{4\bar R\bar\omega_{IAS}}\ll 1$ smaller than the absolute value 
of the r.h.s in Eq.~(\ref{MFappr}).
Thus, by subtracting $\hat{\cal V}_C^t$ from $\hat V_C^t$ in Eq.~(\ref{nuPotF}) only a small change in $M^{0\nu}_F$ is introduced. Such a subtraction with an appropriate choice of $\bar R \sim R$ allows to cut off the contribution to $M^{0\nu}_F$ from the long internucleon distances, where $\frac{1}{r_{ab}}$ has a smooth behavior and which are relevant for the Coulomb mean field.
Therefore, the major contribution to $M^{0\nu}_F$ should come from the short distances 
where the gradient of $\frac{1}{r_{ab}}$ is the largest. 
This provides a natural qualitative explanation of the numerical results of both the QRPA and NSM~\cite{Rod03a,cau96} 
which consistently show the short-range character of the partial $r$-dependent contribution to $M^{0\nu}$.

Of course, by measuring only $M_{F}^{0\nu}$ one does not get the total m.e. $M^{0\nu}$ but rather its 
subleading contribution. However, knowledge of $M_{F}^{0\nu}$ itself brings a very important piece of information. For instance, it will allow to investigate the $A$ dependence of $M_{F}^{0\nu}$. Also, it can help a lot to discriminate between  
different nuclear structure models in which calculated $M_{F}^{0\nu}$ may differ by as much as a factor of 5.
In addition, the ratio $M_{F}^{0\nu}/M_{GT}^{0\nu}$ may be more 
reliably calculable in different models than $M_{F}^{0\nu}$ and $M_{GT}^{0\nu}$ separately. 
Let us put forward here some simple arguments in support of the latter statement. 
Since only small internucleon distances determine $M^{0\nu}$, 
then only nucleon pairs in the spatial relative $s$-wave must dominantly contribute to the m.e.. 
The isotensor Coulomb interaction only couples $T=1$ pairs which must then be in the state with the total spin $S=0$ to assure antisymmetry of the total two-body wave function. 
Because of this and the fact that $\sss_1 \cdot \sss_2|S=0,T=1 \rangle = -3 |S=0,T=1 \rangle$, a natural estimate for the Gamow-Teller m.e. is 
$M^{0\nu}_{GT}=-3 M^{0\nu}_F$ 
provided the neutrino potential is the same in both F and GT cases.
The high-order terms of the nucleon weak current which are present in the case of the GT m.e., but absent in the F m.e.,
change a bit this simple estimate to $M_{GT}^{0\nu}/M_{F}^{0\nu}\approx -2.5$. Also, an uncertainty of few percent 
may come from the difference in the mean nuclear excitation energies in the F and GT cases. 
It is worth noting that the recent QRPA results~\cite{Rod03a,Rod08,Kort07} 
are in good correspondence with these simple estimates.

Here, we want to estimate possible corrections to the simplest closure approximation discussed above. Due to universality and conservation of the vector current, all the corrections of the vector current vertices should be the same independently of which virtual particle, neutrino or photon, is exchanged between them. This is true for the effects of short-range correlations and the finite nucleon size. A small difference of a few procent in the realistic potentials may arise from different mean nuclear excitation energies while exchanging the neutrino or photon but this effect seems 
to be rather reliably calculable.
Another difference can arise from those corrections to the propagator of the virtual photon, as for instance the vacuum polarization correction, that are missing in the case of the virtual neutrino. The effect of the the vacuum polarization is about 0.5\%
and can simply be accounted for by a proper renormalization of the electron charge.

The effect of isospin nonconservation in the strong two-body interaction can be estimated to be at the level of 
2\%--3\%~\cite{auer72,auer83}. One can then directly compare the radial dependencies of the isospin-breaking part of the two-body strong interaction in the $S=0,T=1$ channel and 
the Coulomb interaction within the relevant short range of 1--2 fm to find the dominating source 
of the isospin breaking. Following Ref.~\cite{Eric88}, one can approximate the radial dependence of 
the isospin-breaking strong two-body central potential as (0.02--0.03)$\times\frac{f_\pi^2}{4\pi}\frac{e^{-m_\pi r}}{r}$ ($\hbar= c=1$). 
With $\frac{f_\pi^2}{4\pi}\approx 0.08$ one arrives at the conclusion that this source of the isospin 
non-conservation must be about 20--30 \% of that caused by the Coulomb interaction. 
Though there are rather large relative uncertainties in calculating the isospin-breaking part of 
the two-body strong interaction, by assuming that this correction could in principle be evaluated 
with a moderate accuracy of 30 \%, a residual uncertainty of only 10 \% in $M^{0\nu}_F$ is thereby induced.

Thus, the main message of this Rapid Communication that, at least in principle, $M^{0\nu}_F$ is measurable remains intact in the most realistic situation (though minor corrections may be needed).

To conclude, we have shown in this Rapid Communication that the Fermi $0\nu\beta\beta$ nuclear m.e. can be reconstructed 
if one is able to measure the isospin-forbidden Fermi m.e. between the ground state of the final nucleus and the isobaric analog state in the intermediate nucleus, for instance by means of charge-exchange reactions of the $(n,p)$-type. 
Knowledge of $M_{F}^{0\nu}$ would bring a quite important piece of information on the total $0\nu\beta\beta$ nuclear m.e.. Simple arguments show that the estimate $M_{GT}^{0\nu}/M_{F}^{0\nu}\approx -2.5$ should hold. Also, 
such a measurement can help to discriminate between different nuclear structure models 
in which calculated $M_{F}^{0\nu}$ may differ by as much as a factor of 5.

The authors acknowledge support of both the Deutsche Forschungsgemeinschaft within the SFB TR27 ``Neutrinos and Beyond'' and the EU ILIAS project under Contract No. RII3-CT-2004-506222.


\begin{thebibliography}{99}
\bibitem{Kay89} B.~Kayser, F.~Gibrat-Debu, and F.~Perrier, {\it The Physics of Massive Neutrinos} (World Scientific, Singapore, 1989).
%
\bibitem{Moh91} R.N. Mohapatra and P.B. Pal, {\it Massive Neutrinos in Physics and Astrophysics} (World Scientific, Singapore, 1991).
%
\bibitem{Moh04} R.~N.~Mohapatra {\it et al.}
arXiv:hep-ph/0412099;
R.~N.~Mohapatra and A.~Y.~Smirnov,
  Ann.\ Rev.\ Nucl.\ Part.\ Sci.\  {\bf 56}, 569 (2006).
\bibitem{see-saw}
M.~Gell-Mann, P.~Ramond and R.~Slansky,
in \textit{Supergravity}, edited by F. van Nieuwenhuizen and D.
  Freedman (North Holland, Amsterdam, 1979), p.~315;
T.~Yanagida, {\it Proceedings of the Workshop on Unified Theory and the Baryon Number of the
  Universe} (KEK, Japan, 1979); 
P. Minkovski, Phys. Lett. {\bf B67}, 421 (1977);
S.L. Glashow, {\it Proceedings of Cargese Summer Institute on Quarks and Leptons} (Plenum Press, 
New York, 1980), pp. 687-713;
R.N. Mohapatra and G.~Senjanovi{\'c}, Phys. Rev. Lett. \textbf{44}, 912 (1980).

\bibitem{Kay08}
  B.~Kayser,
  arXiv:0804.1497 [hep-ph], pp. 163-171 in 
 ``Review of particle physics'', Phys.\ Lett.\  B {\bf 667}, 1 (2008);
C.~Giunti and C.~W.~Kim,
  {\it  Fundamentals of Neutrino Physics and Astrophysics}
(University Press, Oxford, UK, 2007);
S.M. Bilenky, C. Giunti, J.A. Grifols, E. Mass\'o,
  Phys. Rep. {\bf 379}, 69 (2003);
P. Langacker, Int. J. Mod. Phys. {\bf A20}, 5254 (2005);
S. J. Freedman and B. Kayser, physics/0411216; 
A. Strumia and F. Vissani, hep-ph/0606054;
R.~D.~McKeown and P.~Vogel, Phys. Rep. \textbf{394}, 315 (2004);
G. L. Fogli, E. Lisi, A. Marrone and A. Palazzo,
Progr. Part. Nucl. Phys. {\bf 57}, 742 (2006);
  T.~Schwetz, M.~Tortola and J.~W.~F.~Valle,
  New J.\ Phys.\  {\bf 10}, 113011 (2008).
%
\bibitem{SV82} J. Schechter and J.W.F. Valle, Phys. Rev. D {\bf 25}, 774 (1982).
\bibitem{vogelbook} F. Boehm and P. Vogel, {\it Physics of Massive Neutrinos}, 2nd
ed. (Cambridge University Press, Cambridge, 1992).
\bibitem{fae98} A.~Faessler and F.~\v{S}imkovic, J. Phys. G {\bf 24}, 2139 (1998);
J.~Suhonen and O.~Civitarese, Phys. Rep. {\bf 300}, 123 (1998);
S.R.~ Elliott and P.~Vogel, Annu. Rev. Nucl. Part. Sci. {\bf 52}, 115 (2002);
J.D. Vergados, Phys. Rep. {\bf 361}, 1 (2002);
S.~R.~Elliott and J.~Engel, J.\ Phys.\ G {\bf 30}, R183 (2004);
C. Aalseth et al., hep-ph/0412300.
\bibitem{AEE07} Frank T. Avignone III, Steven R. Elliott, and Jonathan Engel,  Rev. Mod. Phys. {\bf 80}, 481 (2008). 
%
%
\bibitem{Rod03a} V.A.~Rodin, A.~Faessler, F.~\v Simkovic and P.~Vogel, Phys.\ Rev.\ C {\bf 68}, 044302 (2003);
V.~A.~Rodin, A.~Faessler, F.~Simkovic and P.~Vogel, 
Nucl.\ Phys.\  {\bf A766}, 107 (2006);
{\it ibid.} {\bf A793}, 213(E) (2007);
F.~\v Simkovic, A.~Faessler, V.A.~Rodin, P.~Vogel, and  J. Engel, Phys. Rev. C {\bf 77}, 045503 (2008);
F.~\v Simkovic, A.~Faessler, H.~M\"uther, V.A.~Rodin, and M.~Stauf, Phys. Rev. C {\bf 79},  055501 (2009).

\bibitem{Kort07} M. Kortelainen, O. Civitarese, J. Suhonen, and J. Toivanen,
Phys. Lett. {\bf B647}, 128 (2007); 
M. Kortelainen and J. Suhonen, Phys. Rev. C {\bf 75}, 051303(R) (2007);
  M.~Kortelainen and J.~Suhonen,
  Phys.\ Rev.\  C {\bf 76}, 024315 (2007); 
  J.~Suhonen and O.~Civitarese,
  Phys.\ Lett.\  B {\bf 668}, 277 (2008).
\bibitem{Rod08} V.~Rodin and A.~Faessler,
  Phys.\ Rev.\  C {\bf 77}, 025502 (2008).

\bibitem{cau96}   E. Caurier, F. Nowacki, A. Poves, J. Retamosa,
            Phys. Rev. Lett. {\bf 77}, 1954 (1996);
  E. Caurier, F. Nowacki, A. Poves, Eur. Phys. J. A {\bf 36}, 195 (2008);
  E. Caurier, J. Men\'endez, F. Nowacki, A. Poves, Phys. Rev. Lett. {\bf 100}
  052503 (2008);
J. Men\'endez, A. Poves, E. Caurier, F. Nowacki,
Nucl. Phys. A  {\bf 818}, 139 (2009).
\bibitem{frekers}
  S.~Rakers {\it et al.},
  Phys.\ Rev.\  C {\bf 70}, 054302 (2004);
  S.~Rakers {\it et al.},
  Phys.\ Rev.\  C {\bf 71}, 054313 (2005);
  D.~Frekers,
  Prog.\ Part.\ Nucl.\ Phys.\  {\bf 57}, 217 (2006);
  E.~W.~Grewe {\it et al.},
  Phys.\ Rev.\  C {\bf 76}, 054307 (2007);  
  E.~W.~Grewe {\it et al.},
  Phys.\ Rev.\  C {\bf 78}, 044301 (2008);
  H.~Dohmann {\it et al.},
  Phys.\ Rev.\  C {\bf 78}, 041602 (R) (2008).
\bibitem{auer72} N.~Auerbach, J.~H\"ufner, A.~K.~Kerman and C.~M.~Shakin,
  Rev.\ Mod.\ Phys.\  {\bf 44}, 48 (1972).
\bibitem{auer83} N.~Auerbach, Phys.\ Reps. {\bf 98}, 273 (1983).
\bibitem{Rodin05} V.A.~Rodin, M.H.~Urin, A.~Faessler, Nucl. Phys. A {\bf 747}, 297 (2005).
\bibitem{Rob95} D.A.~Roberts et al., Phys. Rev. C {\bf 52}, 1361 (1995).
\bibitem{Eric88} T.~E.~O.~Ericson and W.~Weise,  {\it  Pions and Nuclei} (Clarendon, Oxford, UK, 1988).
\end{thebibliography}
\end{document}